# TAYLOR'S POWER LAW: BEFORE AND AFTER 50 YEARS OF SCIENTIFIC SCRUTITY

XU MENG[1,*]

[1] *Department of Mathematics and Physics, University of New Haven, 300 Boston Post Road, West Haven, CT 06516. United States of America*

*\*Corresponding author*
*Email: mxu@newhaven.edu*

**Abstract**. Taylor's power law is one of the mostly widely known empirical patterns in ecology discovered in the 20th century. It states that the variance of species population density scales as a power-law function of the mean population density. Taylor's power law was named after the British ecologist Lionel Roy Taylor. During the past half-century, Taylor's power law was confirmed for thousands of biological species and even for non-biological quantities. Numerous theories and models have been proposed to explain the mechanisms of Taylor's power law. However an understanding of the historical origin of this ubiquitous scaling pattern is lacking. This work reviews two research aspects that are fundamental to the discovery of Taylor's power law and provides an outlook of its future studies.

**Introduction**

Lionel Roy Taylor (1914-2007), a British ecologist, published an article entitled "Aggregation, variance and the mean" in the March 4, 1961 issue of *Nature*. In the paper, Taylor studied the relationship between the mean and the variance of population density for at least 22 biological species in 24 data sets (Taylor 1961, referred to as *TL61* hereafter). On the log-log scale, Taylor showed that the mean-variance relationship was described well by a linear equation

$$\log(variance) = \log a + b \log(mean), a > 0. \quad (\text{eqn 1})$$

On arithmetic scale, eqn 1 becomes a power law

$$variance = a(mean)^b, a > 0. \quad (\text{eqn 2})$$

Eqn 2 earned the name "Taylor's law" (not to be confused with "Taylor law", the Public Employees Fair Employment Act) or "Taylor's power law (TPL)" after extensive empirical confirmations using aphid, moth, and bird data collected in England by Taylor and his colleagues (Taylor, Woiwod, Perry 1978; Taylor, Woiwod 1980; Taylor, Woiwod 1982; Taylor, Taylor, Woiwod, Perry 1983; Taylor 1984). The exact phrase "Taylor's power law" first appeared in the paper by D. G. Harcourt (1963), where the author studied the spatial distribution of Colorado beetle, a pest in potato crops. In the recent decades, studies on TPL were reinvigorated by the increased power of statistical computation and improved accessibility to ecological data sets (Brown et al. 2002; Marquet et al. 2005; Eisler, Bartos, Kertész 2008). *TL61*, widely perceived as the founding work of TPL, has been cited in roughly 2174 literature up to today (last retrieved from Google Scholar on 04/07/2016).

In the past half-century, both theoretical and applied studies were conducted on TPL. Numerous biological and statistical models have been proposed to explain the underlying mechanisms of TPL (Taylor, Taylor 1977; Hanski 1987; Perry 1988; Kilpatrick, Ives 2003; Cohen, Xu, Schuster 2013; Giometto et al. 2015; Cohen, Xu 2015; Xiao et al. 2015). A list of physical models relevant to TPL can be found in Eisler et al. (2008). In practice, TPL has been applied to the sampling design of agricultural pest (Green 1970; Kuno 1991; Celini, Vaillant 2004), sample size estimation of fish species (Mouillot et al. 1999), and quantification of temporal variability in seabird (Certain et al. 2007) and reef fish (Mellin et al. 2010) populations.

TPL was debated on its mathematical formulation and ecological implications. Several competing models of TPL have been proposed based on their statistical superiority in data fitting (Routledge, Swartz 1991; Perry, Woiwod 1992; Tokeshi 1995). Other scientists questioned TPL's biological interpretation and usefulness (Anderson et al. 1982; Downing 1986; Kuno 1991). These critiques propelled the theoretical development of TPL, however a convincing and unified theory of TPL is still lacking. Until such a theory can be found or the contradicting evidences of TPL (Anderson et al. 1982; Downing 1986) can be validated, understanding of this ubiquitous pattern remains incomplete.

This paper reviews the historical developments of mean-variance scaling before and after TL61, and highlights several unsolved research questions of TPL. To the author's knowledge, the history of TPL has not been thoroughly investigated in the existing literature. For modern ecologists, testing and modeling empirical patterns have become a major component of their work, while it is equally important to appreciate and beware of the scientific roots of established concepts and models. Through this retrospective thinking the limitation of examined patterns can be exposed, a prudent model-building approach can be taken, and unsolved open questions can be revealed. It is the hope of this work that it will raise the awareness of the historical developments of TPL among quantitative ecologists, and aid their research on this ubiquitous empirical pattern.

I review the research of mean-variance scaling before *TL61* and TPL after *TL61* in sections 2 and 3 respectively. In section 4 I present a statistical model that is conceptually equivalent to but does not explain TPL. In section 5 I discuss some key issues that have been overlooked in the research of TPL.

**Mean-variance scaling before TL61**

Studies on the mean-variance scaling began even before the term "variance" was invented (Fisher 1918). In his 1879 article on thermal aerodynamics (Reynolds 1879), the famous physicist Osborne Reynolds plotted the pressure difference against pressure on the log-log axes, and compared the difference between hydrogen and air. Had Reynolds used variance to quantify pressure differences, the origin of mean-variance scaling would be traced back before any statistician or biologist started pursuing this topic. The theory underlying Reynolds' plot (Fig. 1a) is beyond the expertise of the author and therefore not described here.

In statistics, the correlation between the sample mean and higher-order sample moments was derived in an editorial work of *Biometrika* (author unknown 1903, pp. 279), although the same result had been observed by Pearson and Filon (1897, pp. 236). Neyman (1926, pp. 402) derived the regression coefficient between the sample variance (dependent variable) and the sample mean (independent variable).

Mean-variance relationship gained broad scientific attentions when analysis of variance (ANOVA, Fisher 1918, 1921) was developed and applied in natural sciences. One of the key assumptions in ANOVA was the equality of variances across different experimental groups (homoscedasticity). However, for distributions of most real-world data, variance is not a constant but varies with the mean. To apply ANOVA, scientists studied mean-variance scaling and designed variable transformations that can stabilize the variance. For example, Bartlett (1936a) applied squared root transformation on Poisson random variable to derive a constant variance. In a separate work, Bartlett (1936b) proposed an analytic relationship between the sample mean and the sample variance:

$$sample\ variance = \alpha(sample\ mean) + \beta(sample\ mean)^2. \qquad \text{(eqn 3)}$$

Bartlett used eqn 3 to quantify variance of crane fly (*Nephrotoma appendiculata*) larvae populations, and showed that it provided a better fit than Poisson, which leads to mean = variance. Fisher et al. (1921) also used eqn 3 to describe bacterial population densities. Clapham (1936) found that distributions of individual plants in prairie were not Poisson but "over-dispersed" (variance > mean). Five years later, Bliss (1941) used the log-linear form of TPL (eqn 1) to fit spatially grouped populations of Japanese beetle larvae. Bliss's article was the first to publish the current form of TPL and pushed the origin of TPL 20 years before Taylor's work in 1961[1]. Later, Beall (1942) systematically studied the mean-variance relationship. He noticed that in entomological field data, "the departure of $s^2$ (sample variance) from $\bar{x}$ (sample mean) becomes disproportionately great as $\bar{x}$ increases", and used the functional mean-variance relationship of a negative binomial distribution to account for this discrepancy. Namely,

$$variance = mean + k(mean)^2. \qquad (eqn\ 4)$$

Fracker and Brischle (1944) pointed out that eqn 4 was inadequate to describe ribes populations because $k$ changed with the mean and the quadrat sizes. They found Bartlett's equation (eqn 3) and TPL (eqn 2) yielded better fits.

Entering 1950's, researchers continued to analyze the mean-variance scaling and to explore different variance stabilizing transformations of interested variables, such as tasters' scores on food (Hopkins 1950) and drink (Coote 1956), and catches of marine species (Barnes 1952). Taylor (1961), using the power-law relationship (eqn 2) tested for 24 ecological population data sets, defined a transformation that stabilizes the variance as a constant (see eqn (3) in *TL61*). Unfortunately, the transformation formula in *TL61* was incomprehensible because Taylor did not provide any mathematical derivations. Below I included a short proof of Taylor's formula.

For a random variable *X*, a variance-stabilizing function *f* is defined such that, for a constant *Q*,

$$s^2(f(X)) = Q.$$

Using the first-order delta method (Oehlet 1991),

$$Q = s^2(f(X)) \approx s^2(X) \cdot [f'(m)]^2.$$

If TPL held with parameters *a* and *b* as specified in eqn 2, then above equation becomes

$$am^b \cdot [f'(m)]^2 = Q,$$

or

---

[1] Eisler et al. (2008) reported that TPL was first discovered by H. Fairfield Smith (1938). However a re-examination of Smith's paper showed that the author did not study the relationship between variance of yield of agricultural crops and the corresponding mean yield, but the relationship between the variance of yield and plot size.

$$f'(m) = \left(\frac{Q}{a}\right)^{\frac{1}{2}} m^{-\frac{b}{2}}.$$

Integrating both sides of the above equation yields

$$f(m) = \left(\frac{Q}{a}\right)^{\frac{1}{2}} \int_t^m u^{-\frac{b}{2}} du. \qquad \text{(eqn 5)}$$

Eqn 5 is identical to eqn (3) in *TL61*. The form of *f* depends on the power exponent *b* of TPL and *t*. For example, when *b* = 1 and *t* ≥ 0, *f* is a square root function. When *b* = 2 and *t* > 0, *f* is a natural logarithmic function. When *b* = 4, *t* ≠ 0, *f(x)* is a linear function of 1/*x*.

**Taylor's power law after TL61**

The significance of TL61 does not necessarily lie in its discovery of the power-law mean and variance scaling of species population densities, but in that it is the first meta-analysis confirming TPL and establishing it as one of the quantitative patterns in ecology (Smith et al. 2014). In the past half-century, Taylor's pioneer work has inspired many biologists to test TPL against thousands of biological taxa. Such examples can be found in the review by Eisler et al. (2008). As more and more empirical support of TPL was found, the scientific interest of TPL has shifted from its empirical confirmation to its underlying mechanisms. In this section, I will briefly discuss the theoretical development of TPL in the past few decades, especially on the biological and statistical models of TPL.

Three important works addressed the biological mechanisms of TPL from different perspectives. Taylor and Taylor (1977) used density-dependent power law functions that model animal migratory behaviors and account for TPL. Anderson et al. (1982), on the other hand, illustrated TPL using classic population models with demographic and environmental stochasticity, without incorporating any behavioral mechanism. Based on a stochastic logistic model of multiple species, Kilpatrick and Ives (2003) showed that increased interspecific competition reduced the slope of TPL to less than two. Several other authors (Perry 1988, 1994; Ballantyne 2005) also have used population dynamic models to explain TPL and its parameter values.

Recently, the statistical reasons of TPL have gained considerable attention, because the large variety of taxa and models confirming TPL indicated that a unified mechanism-independent theory must at work (Cohen and Xu 2015). Among several explanations are the skewed distribution theory (Cohen and Xu 2015), plausible set theory (Xiao et al. 2015), and large deviation theory (Giometto et al. 2015). The common theme in these works is that the appearance of TPL does not rely on specific biological processes. This school of thoughts provides new perspectives in tackling TPL's ubiquity in nature.

**Tweedie's distributions and TPL**

15 years before *TL61*, the British statistician, Maurice C.K. Tweedie (1946), asked the question: "How can one determine the types of distribution in which the regression function (sample variance as a function of sample mean) has a specified polynomial form?" Tweedie was the first to explore the relationship between underlying probability distribution and mean-variance scaling (when the polynomial function becomes a power law). An important contribution by Tweedie was that he defined an "exponential family (or Laplacian distribution)" for which the variance can be written as a polynomial function of the mean (Tweedie 1947). A similar study when the variance was a quadratic function of the mean was conducted by Morris (1982). Tweedie (1984) gave special attention to a class of Laplacian distribution $F(\alpha)$ such that

$$\frac{d(\ln k_2)}{d(\ln k_1)}$$

is a constant independent of the distribution parameter $\alpha$. Here $k_1$ and $k_2$ are the first and second cumulants of the distribution respectively. Using the definition of cumulants, the above quantity becomes

$$F''(\alpha) = a\bigl(-F'(\alpha)\bigr)^b$$

Interestingly, this equation is equivalent to the form of TPL (eqn 2) and has the solution

$$F(\alpha) = \begin{cases} \frac{1}{a(-1)^b(2-b)}\left[\bigl(a(-1)^b(1-b)\alpha + c_1^{1-b}\bigr)^{\frac{2-b}{1-b}} - c_1^{2-b}\right] + c_2 & \text{for } b \neq 1 \text{ or } 2 \\ \frac{c_1}{a}(1 - e^{-a\alpha}) + c_2 & \text{for } b = 1 \\ -\frac{\log(a - c_1 a\alpha)}{a} + c_2 & \text{for } b = 2 \end{cases}$$

where $c_1 = F'(0)$ and $c_2 = F(0)$ (Tweedie 1984, pp. 582). Jørgensen (1997) gave a mathematically equivalent solution. To summarize, Tweedie found explicitly a family of distribution functions that satisfied TPL.

Kendal and colleagues (2004, 2011) argued that Tweedie's model provides a universal explanation of TPL and defined a family of distributions that satisfy TPL, "the Tweedie distributions". Several fundamental caveats exist in their theory. First, Kendal et al. failed to notice that Tweedie's approach works for a polynomial relationship between variance and mean, not only for the mean-variance power law described by TPL. Second, in all claims made by Kendal and his colleagues, TPL (or "invariance under scale transformation") was assumed a priori as a given assumption (Kendal 2004, pp. 202; Kendal, Jørgensen 2011, pp066115-2, 3). This misuse of TPL by assuming its validity, instead of deriving it, disproves the Tweedie's model as an explanation of TPL. Third, as pointed out by Cohen et al. (2013), Tweedie distributions do not explain TPL with power exponent between 0 and 1, which were found in some empirical studies (Green 1970; Keil et al. 2010, Fig. 2). Overall, "Tweedie distributions" or "Tweedie's model" do not provide a universal explanation of TPL, but merely incorporated a class of well-known probability distributions for which the variance can be written as a power-law of the mean, with a restricted range in the power-law exponent (see Table 1 in Kendal 2004, pp. 203).

**Conclusions and outlooks of TPL research**

My review showed that the development and application of ANOVA set up the theoretical framework and necessity for the discovery of TPL. TPL was neither universal nor superior to other mean-variance scaling relationships (Downing 1986; Routledge and Swartz 1991) in the history. The meaning and usefulness of TPL rely on confirmation against empirical data, which in turn requires new ways of data synthesis and analysis.

As stated in the **Introduction**, one of the challenges in the research of TPL is that no unified theory or model has yet been found. Most existing models of TPL relied on the biological features of particular species or specific environmental and experimental conditions, and lacked generality to be useful for various species across multiple scales. Such problem results from the limited scope and scale of population abundance data used in the testing of TPL. Taylor and colleagues confirmed TPL using aphid, insect and bird population data collected from the Rothamsted Experimental Station (currently Rothamsted Research) and throughout Great Britain (see Fig. 2 in Taylor, Taylor 1977 and Fig. 1 in Taylor, Woiwod 1980). However the species and geographical range examined by Taylor may be limited to prove the universality of TPL. A synthetic analysis of TPL using multiple large-scale data sets is therefore necessary to reveal important properties of TPL (e.g. scale dependence, species differences) that may be unobservable at a local scale. The following sections describe three current issues in the studies of TPL and elaborate how multi-scale data analysis will help resolve these issues.

*Scales in TPL*

If an empirical pattern depended on specific temporal or spatial scales on which it was tested, then the generality of this pattern and its usefulness should be questioned. The testing of TPL relied on three scale measures: First, the size of quadrat or sampling site, which is the area of habitat that the species lives and determines how many individuals are included to compute the population count. Second, the number of quadrats or sites within a block (or a super-cluster of quadrats or sites), or sample size, which affects the accuracy in the estimations of the mean and the variance. Third, the number of blocks used in the statistical fit of TPL, which reveals the overall scale of the study site. All three scale measures contain information on the size and location of spatial units, and their changes will lead to statistical or biological consequences to the estimation of TPL parameters.

To examine the relation between scales and TPL, a naïve combination of data from multiple experiments at various scales will not work, since the observed scale effects are likely to be confounded with specific experimental methods or environmental conditions. Existing literature studied the effect of quadrat size (Sawyer 1989) and sample size (Clark, Perry 1994) on TPL and its parameters (slope and intercept in eqn 3). However in both works the conclusions were based on simulated population counts instead of real empirical data, and therefore cannot be applied to realistic ecological scenarios. In fact, the main reason Sawyer (1989) used a simulation approach, as he claimed, was the limitations of multiple scales in the ecological data sets. Such caveat can be overcome with the use of multi-scale data that are currently available (e.g. Breeding Bird Survey, Forest Inventory Analysis, North America Butterfly Association).

*Estimation of mean and variance*

The sample mean and sample variance of a species population exhibit interesting self-restraint properties that may affect the parameters of TPL. For examples, in a sample of population counts of size $n$, the sample variance is limited between 0 and $n\bar{x}^2$ (here $\bar{x}$ being the sample mean, see Tokeshi 1995). Another issue in the testing of TPL is the underestimation of population variance (and of population mean, to a less extent) of skewed distributions (Ross 1990), which are often observed in species population counts. Underestimated mean or variance will distort the true behavior of TPL using population mean and variance, and yield uninterpretable statistical artifacts. While the actual impact of this phenomenon on TPL and its parameters remains to be seen, ecological data sets of large sample sizes may mitigate this issue to some extent.

*Species specificity*

Are the parameters of TPL specific to species? The most important work that addressed this question was written by Downing (1986), where the author showed that the values of exponent $b$ of TPL may be similar among different species but vary according to environments. Using published data, Downing also showed that the size of scale measures (see *Scales in TPL*) affected the values of $b$, casting doubt to the applicability of TPL. Taylor and colleagues (1988) disputed Downing's finding on its statistical method and data quality, but did not provide a strategy to examine the species specificity of the parameters of TPL. Meta-analysis of population data with comparable scale measures could potentially answer this fundamental question.

**Literature Search**

Literature on mean-variance scaling and the use of logarithm in bivariate studies were searched using exact phrases "variance function", "variance law", "mean-variance", "mean and variance", "analysis of variance", "allometry", "logarithmic transformation", and "logarithmic scale" as the topics of journal articles published before 1961 in the Web of Science database and again in Google Scholar. Same phrases were also searched in the archives of three oldest statistical journals in the world: *Journal of the Royal Statistical Society* (first issue in 1838), *Journal of the American Statistical Association* (first issue in 1888), and *Biometrika* (first issue in 1901) for relevant literature. Cross-referencing was conducted in relevant articles.

**Acknowledgements**

The author thanks Joel E. Cohen for constructive comments and suggestions on an earlier draft of the manuscript. The research was supported by University of New Haven Summer Research Fund and NSF Grant No. 1038337 to the Rockefeller University.

**References**

1. Anderson, R.M., Gordon, D.M., Crawley, M.J., Hassell, M.P. (1982): Variability in the abundance of animal and plant species. – Nature 296: 245–248.
2. Ballantyne IV, F. (2005): The upper limit for the exponent of Taylor's power law is a consequence of deterministic population growth. – Evolutionary Ecology Research 7: 1213–1220.

3. Barnes, H. (1952): The use of transformations in marine biological statistics. – Journal du conseil 18: 61–71.
4. Bartlett, M.S. (1936a): The square root transformation in analysis of variance. – Supplement to the Journal of the Royal Statistical Society 3: 68–78.
5. Bartlett, M.S. (1936b): Some notes on insecticide tests in the laboratory and in the field. – Supplement to the Journal of the Royal Statistical Society 3: 185–194.
6. Beall, G. (1942): The transformation of data from entomological field experiments so that the analysis of variance becomes applicable. – Biometrika 32: 243–262.
7. Bliss, C.I. (1941): Statistical problems in estimating populations of Japanese beetle larvae. – Journal of Economic Entomology 34: 221-232.
8. Brown, J.H., Gupta, V.K., Li, B-L., Milne, B.T., Restrepo, C., West, G.B. (2002): The fractal nature of nature: power laws, ecological complexity and biodiversity. – Philosophical Transactions of the Royal Society of London B 357: 619–626.
9. Celini, L., Vaillant, J. (2004): A model of temporal distribution of Aphis gossypii (Glover) (Hem., Aphididae) on cotton. – Journal of Applied Entomology 128: 133–139.
10. Certain, G., Bellier, E., Planque, B., Bretagnolle, V. (2007): Characterising the temporal variability of the spatial distribution of animals: an application to seabirds at sea. – Ecography 30: 695–708.
11. Clapham, A.R. (1936): Over-dispersion in grassland communities and the use of statistical methods in plant ecology. – Journal of Ecology 24: 232–251.
12. Clark, S.J., Perry, J.N. (1994): Small sample estimation for Taylor's power law. – Environmental and Ecological Statistics 1: 287–302.
13. Cohen, J.E., Xu, M., Schuster, W.S.F. (2013): Stochastic multiplicative population growth predicts and interprets Taylor's power law of fluctuation scaling. – Proceedings of the Royal Society B 280: 20122955.
14. Cohen, J.E., Xu, M. (2015): Random sampling of skewed distributions implies Taylor's power law of fluctuating scaling. – Proceedings of the National Academy of Sciences U.S.A. 112: 7749–7754.
15. Coote, G.G. (1956): Analysis of scores for bitterness of orange juice. – Journal of Food Science 21: 1–10.
16. Downing, J.A. (1986): Spatial heterogeneity: evolved behaviour or mathematical artefact? – Nature 323: 255–257.
17. Editorial (1903): On the probable errors of frequency constants. – Biometrika 2: 273–281.
18. Eisler, Z., Bartos, I., Kertész, J. (2008): Fluctuation scaling in complex systems: Taylor's law and beyond. – Advances in Physics 57: 89–142.
19. Fisher, R.A. (1918): The correlation between relatives on the supposition of Mendelian inheritance. – Philosophical Transactions of the Royal Society of Edinburgh 52: 399–433.
20. Fisher, R.A. (1921): Studies in crop variation. I. An examination of the yield of dressed grain from Broadbalk. – Journal of Agricultural Science 11: 107–135.
21. Fisher, A. (1930): The Mathematical Theory of Probabilities, 2nd edition. – MacMillan Company, New York.
22. Fracker, S.B., Brischle, H.A. (1944): Measuring the local distribution of ribes. – Ecology 25: 283–303.

23. Giometto, A., Formentin, M., Rinaldo, A., Cohen, J.E., Maritan, A. (2015): Sample and population exponents of generalized Taylor's law. – Proceedings of the National Academy of Sciences USA 112: 7755–7760.
24. Green, R.H. (1970): On fixed precision level sequential sampling. – Research on Population Ecology 12: 249–251.
25. Hanski, I. (1987): Cross-correlation in population dynamics and the slope of spatial variance: mean regressions. – Oikos 50: 148–151.
26. Harcourt, D.G. (1963): Population dynamics of Leptinotarsa decemlineata (Say) in Eastern Ontario: I. Spatial pattern and transformation of field counts. – The Canadian Entomologist 95: 813–820.
27. Hopkins, J.W. (1950): A procedure for quantifying subjective appraisals of odor, flavor and texture of foodstuffs. – Biometrics 6: 1–16.
28. Jørgensen, B. (1997): The Theory of Dispersion Models. – Chapman & Hall, London.
29. Kendal, W.S. (2004): Taylor's ecological power law as a consequence of scale invariant exponential dispersion models. – Ecological Complexity 1: 193–209.
30. Kendal, W.S., Jørgensen, B. (2011): Taylor's power law and fluctuation scaling explained by a central-limit-like convergence. – Physical Review E 83: 066115.
31. Keil, P., Herben, T., Rosindell, J., Storch, D. (2010): Predictions of Taylor's power law, density dependence and pink noise from a neutrally modeled time series. – Journal of Theoretical Biology 265: 78–86.
32. Kilpatrick, A.M., Ives, A.R. (2003): Species interactions can explain Taylor's power law for ecological time series. – Nature 422: 65–68.
33. Kuno, E. (1991): Sampling and analysis of insect populations. – Annual Review of Entomology 36: 285–304.
34. Marquet, P.A., Quiñones, R.A., Abades, S., Labra, F., Tognelli, M., Arim, M., Rivadeneira, M. (2005): Scaling and power-laws in ecological systems. – Journal of Experimental Biology 208: 1749–1769.
35. Mellin, C., Huchery, C., Caley, M.J., Meekan, M.G., Bradshaw, C.J.A. (2010): Reef size and isolation determine the temporal stability of coral reef fish populations. – Ecology 91: 3138–3145.
36. Morris, C.N. (1982): Natural exponential families with quadratic variance functions. – The Annals of Statistics 10: 65–80.
37. Mouillot, D., Culioli, J-M., Lepretre, A., Tomasini, J-A. (1999): Dispersion statistics and sample size estimates for three fish species (Symphodus ocellatus, Serranus scriba and Diplodus annularis) in the Lavezzi Islands Marine Reserve (South Corsica, Mediterranean Sea). – P.S.Z.N.: Marine Ecology 20: 19–34.
38. Neyman, J. (1926): On the correlation of the mean and the variance in samples drawn from an "infinite" population. – Biometrika 18: 401–413.
39. Pearson, K., Filon, L.N.G. (1897): On the probable errors of frequency constants and on the influence of random selection on variation and correlation. – Philosophical Transactions of the Royal Society of London A 191: 229–311.
40. Perry, J.N. (1988): Some models for spatial variability of animal species. – Oikos 51: 124–130.
41. Perry, J.N. (1994): Chaotic dynamics generate Taylor's power law. – Proceedings of the Royal Society of London B 257: 221–226.
42. Perry, J.N., Woiwod, I.P. (1992): Fitting Taylor's power law. – Oikos 65: 538–542.

43. Reynolds, O. (1879): On certain dimensional properties of matter in the gaseous state. Part I. Experimental researches on thermal transpiration of gases through porous plates and on the laws of transpiration and impulsion, including an experimental proof that gas is not a continuous plenum. Part II. On an extension of the dynamical theory of gas, which includes the stresses, tangential and normal, caused by a varying condition of gas, and affords an explanation of the phenomena of transpiration and impulsion. – Philosophical Transactions of the Royal Society of London 170: 727–845.
44. Ross, G.J.S. (1990): Incomplete variance functions. – Journal of Applied Statistics 17: 3–8.
45. Routledge, R.D., Swartz, T.M. (1991): Taylor's power law re-examined. – Oikos 60: 107–112.
46. Sawyer, A.J. (1989): Inconstancy of Taylor's b: Simulated sampling with different quadrat sizes and spatial distributions. – Researches on Population Ecology 31: 11–24
47. Smith, H.F. (1938): An empirical law describing heterogeneity in the yields of agricultural crops. – Journal of Agricultural Science 28: 1–23.
48. Smith, F.A., Gittleman, J.L., Brown, J.H. (2014): Foundations of Macroecology. – University of Chicago Press, Chicago.
49. Taylor, L.R. (1961): Aggregation, variance and the mean. – Nature 189: 732–735.
50. Taylor, L.R., Taylor, R.A.J. (1977): Aggregation, migration and population dynamics. – Nature 265: 415–421.
51. Taylor, L.R., Woiwod, I.P., Perry, J.N. (1978): The density-dependence of spatial behaviour and the rarity of randomness. – Journal of Animal Ecology 47: 383–406.
52. Taylor, L.R., Woiwod, I.P. (1980): Temporal stability as a density-dependent species characteristic. – Journal of Animal Ecology 49: 209–224.
53. Taylor, L.R., Woiwod, I.P. (1982): Comparative synoptic dynamics. I. Relationships between inter- and intra-specific spatial and temporal variance/mean population parameters. – Journal of Animal Ecology 51: 879–906.
54. Taylor, L.R., Taylor, R.A.J., Woiwod, I.P., Perry, J.N. (1983): Behavioural dynamics. – Nature 303: 801–804.
55. Taylor, L.R. (1984): Assessing and interpreting the spatial distributions of insect populations. – Annual Review of Entomology 29: 321–357.
56. Taylor, L.R., Perry, J.N., Woiwod, I.P., Taylor, R.A.J. (1988): Specificity of the spatial power-law exponent in ecology and agriculture. – Nature 332: 721–722.
57. Tokeshi, M. (1995): On the mathematical basis of the variance-mean power relationship. – Research on Population Ecology 37: 43–48.
58. Tweedie, M.C.K. (1946): The regression of the sample variance on the sample mean. – Journal of the London Mathematical Society 21: 22–28.
59. Tweedie, M.C.K. (1947): Functions of a statistical variate with given means, with special reference to Laplacian distributions. – Mathematical Proceedings of the Cambridge Philosophical Society 43: 41–49.
60. Tweedie, M.C.K. (1984): An index which distinguishes between some important exponential families. – In: Ghosh, J.K., Roy, J. (Eds.), Statistics: Applications and New Directions. Proceedings of the Indian Statistical Institute Golden Jubilee International Conference, Indian Statistical Institute, Calcutta, India, pp. 579–604.

61. Xiao, X., Locey, K.J., White, E.P. (2015): A process-independent explanation for the general form of Taylor's law. – The American Naturalist 186: E51–E60.

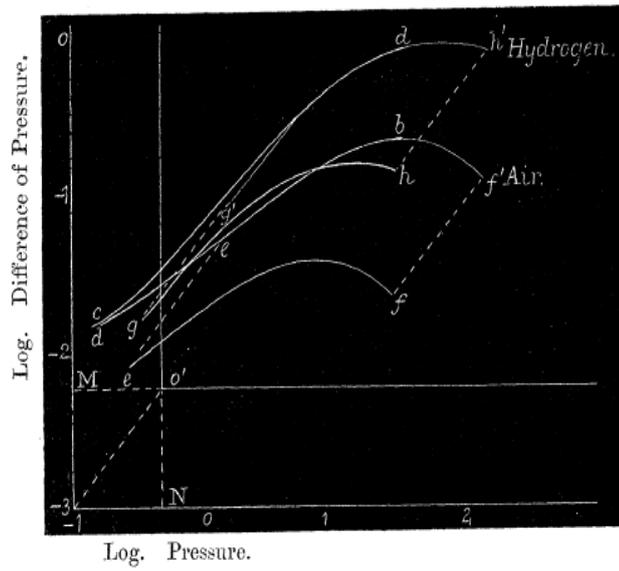

**Figure 1**. The woodcut figure from Reynold (1879) plotting logarithmic difference in pressure against logarithmic pressure.